\documentclass[conference]{IEEEtran}
\IEEEoverridecommandlockouts

\usepackage{cite}
\usepackage{amsmath,amssymb,amsfonts}
\usepackage{graphicx}
\usepackage{textcomp}
\usepackage{xcolor}
\def\BibTeX{{\rm B\kern-.05em{\sc i\kern-.025em b}\kern-.08em
    T\kern-.1667em\lower.7ex\hbox{E}\kern-.125emX}}

\usepackage[utf8]{inputenc} 
\usepackage[T1]{fontenc}    
\usepackage[hidelinks]{hyperref}
\usepackage{url}            
\usepackage{booktabs}       
\usepackage{nicefrac}       
\usepackage{microtype}      
\usepackage{wasysym}

\def\equationautorefname#1#2\null{(#2\null)}

\usepackage{comment}
\usepackage{algorithmicx,algorithm}
\usepackage{algpseudocode}
\usepackage{mathtools}
\usepackage{threeparttable}
\usepackage{multirow}
\usepackage{colortbl} 

\usepackage{caption}
\usepackage{subcaption}
\usepackage{float} 
\begin{document}

\title{ZipVoice: Fast and High-Quality Zero-Shot Text-to-Speech with Flow Matching}

\author{
\IEEEauthorblockN{Han Zhu, Wei Kang, Zengwei Yao, Liyong Guo, Fangjun Kuang, \\
Zhaoqing Li, Weiji Zhuang, Long Lin, Daniel Povey}

\IEEEauthorblockA{
\textit{Xiaomi Corp., Beijing, China}
 \\
\{zhuhan3,dpovey\}@xiaomi.com}
}

\maketitle

\begin{abstract}

Existing large-scale zero-shot text-to-speech (TTS) models deliver high speech quality but suffer from slow inference speeds due to massive parameters. To address this issue, this paper introduces ZipVoice, a high-quality flow-matching-based zero-shot TTS model with a compact model size and fast inference speed. Key designs include: 1) a Zipformer-based vector field estimator to maintain adequate modeling capabilities under constrained size; 2) Average upsampling-based initial speech-text alignment and Zipformer-based text encoder to improve speech intelligibility; 3) A flow distillation method to reduce sampling steps and eliminate the inference overhead associated with classifier-free guidance.
Experiments on 100k hours multilingual datasets show that ZipVoice matches state-of-the-art models in speech quality, while being 3 times smaller and up to 30 times faster than a DiT-based flow-matching baseline. Codes, model checkpoints and demo samples are publicly available.\footnote{\url{https://github.com/k2-fsa/ZipVoice}}
\end{abstract}


\section{Introduction}

Text-to-speech (TTS) aims to synthesize natural speech that accurately reflects the provided text. Zero-shot TTS additionally requires the generated speech to mimic the speaker characteristics of a reference audio, enabling flexible adaptation to arbitrary voices. In recent years, zero-shot TTS systems have witnessed substantial progress, driven by the development of generative modeling methods~\cite{chen2025neural,le2023voicebox,eskimez2024e2,chen2024f5,wang2025maskgct,guo2024fireredtts,anastassiou2024seed} and the availability of large-scale datasets~\cite{zen2019libritts,kang2024libriheavy,he2024emilia}. An ideal TTS system should not only be capable of producing high-quality speech but also maintain simplicity and efficiency. Although existing zero-shot TTS models trained on large-scale datasets have achieved impressive speech quality, they typically rely on a large number of parameters to maintain sufficient modeling capacity. Furthermore, most state-of-the-art (SOTA) zero-shot TTS models employ repeated sampling procedures, such as autoregressive (AR) sampling over time~\cite{chen2025neural} or non-autoregressive (NAR) sampling used in diffusion models~\cite{shennaturalspeech}. The large model size and numerous repeated sampling steps result in slow inference, leading to high deployment costs and limiting the practical applications of zero-shot TTS models.

Recently, flow-matching~\cite{lipman2023flow}, a variant of diffusion models, has been widely adopted in TTS~\cite{le2023voicebox,eskimez2024e2,chen2024f5,li2025sf,kim2023p} due to its reduced sampling steps compared to previous diffusion-based models. A representative flow-matching-based zero-shot TTS system is E2-TTS~\cite{eskimez2024e2}. E2-TTS addresses text-speech alignment by padding text tokens with filler tokens to match the speech length, and letting the Transformer~\cite{vaswani2017attention}-based vector field estimator to learn the alignment implicitly. This approach significantly simplifies the system architecture and training process by obviating the need for token-level duration prediction. 
Nevertheless, achieving satisfactory performance with E2-TTS still necessitates a large model size and dozens of sampling steps, which inevitably results in slow inference.

To address the aforementioned challenges, we introduce ZipVoice, a high-quality zero-shot TTS model with a compact model size and fast inference speed. Built upon the flow-matching framework for its simplicity and superior performance, ZipVoice ensure efficient and high-quality speech generation with following designs. Firstly, Zipformer~\cite{yao2024zipformer} is incorporated as the backbone of vector field estimator to ensure adequate model capacity within a limited parameter budget. We demonstrate that the Zipformer architecture, although originally designed for automatic speech recognition (ASR), is also well-suited for TTS tasks. Secondly, since the practice of padding text tokens with filler tokens in E2-TTS results in suboptimal speech-text alignment and compromised intelligibility~\cite{chen2024f5}, to ensure the intelligibility of ZipVoice, on the one hand, we propose an average upsampling approach that assumes uniform token durations within a sentence. This straightforward strategy preserves system simplicity by eliminating the need for token-level duration prediction, while maintaining stable alignment between speech and text. On the other hand, a Zipformer-based text encoder is employed to extract better text token representations. Third, flow-matching-based TTS models typically require many sampling steps to generate high-quality speech, and the widely used classifier-free guidance (CFG)~\cite{ho2021classifierfree} strategy further introduces an additional inference pass without condition in each sampling step. To mitigate this, we design a flow distillation method to distill the pre-trained flow-matching model, thereby reducing the number of sampling steps and avoiding the extra computational overhead introduced by CFG.

\begin{figure*}[t!]
	\centering
        \begin{subfigure}[b]{0.99\columnwidth}
	\includegraphics[width=\columnwidth]{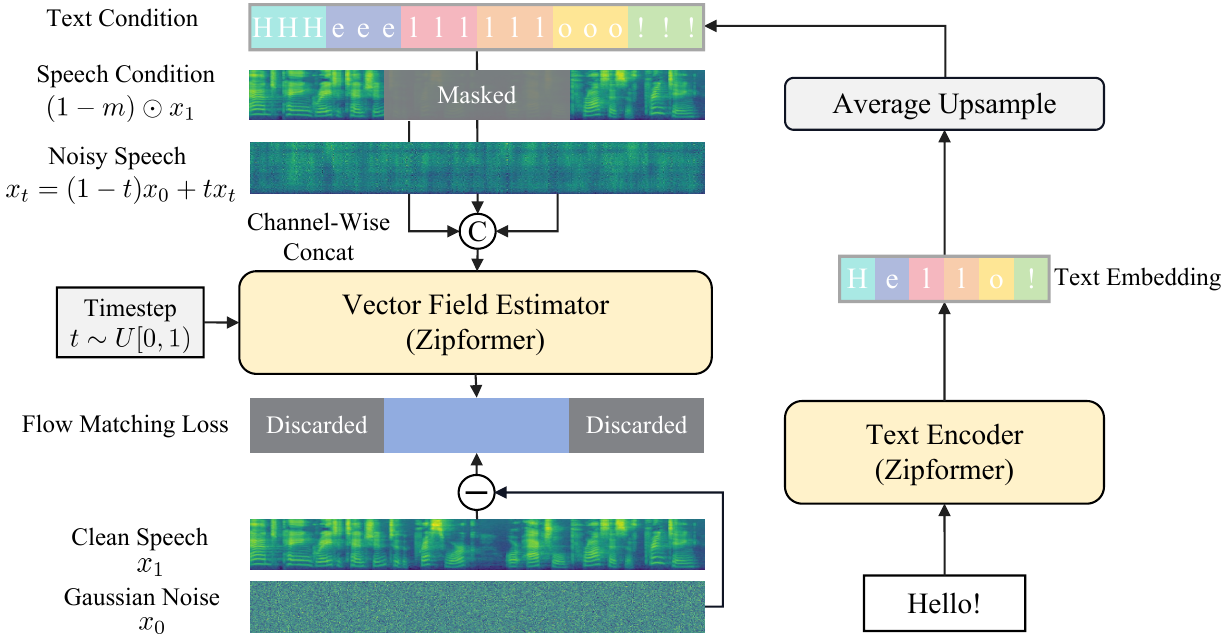}
    \end{subfigure}
        \begin{subfigure}[b]{0.95\columnwidth}
	\includegraphics[width=\columnwidth]{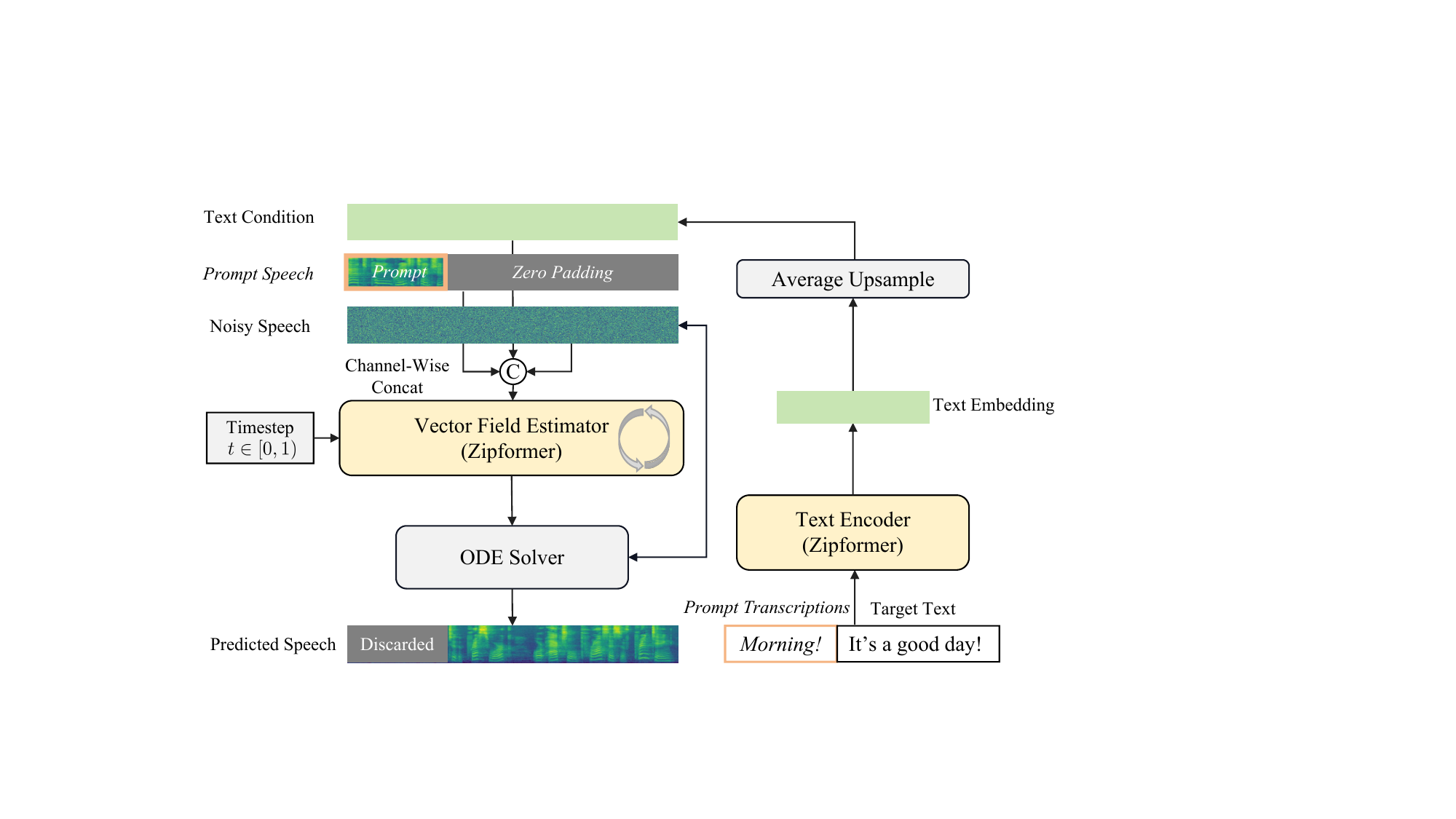}
    \end{subfigure}
	\caption{Illustration of ZipVoice training (left) and inference (right).} 
	\label{fig:framework}
\end{figure*} 

Experimental results demonstrate that, despite having a model size 3 times smaller and an inference speed up to 30 times faster than a DiT-based flow-matching baseline~\cite{chen2024f5}, ZipVoice achieves intelligibility, speaker similarity, and naturalness comparable to existing SOTA zero-shot TTS systems.

\section{ZipVoice}

In this section, we introduce the proposed zero-shot TTS model, ZipVoice. We first outline the flow-matching method as a preliminary. Next, we provide an overview of ZipVoice. Then, we describe its two Zipformer-based sub-models in detail. We further elaborate on the the average upsampling employed in ZipVoice. Subsequently, we present the speedup version of ZipVoice: ZipVoice-Distill, trained with our flow-distillation method. Finally, we detail the inference strategies of the model.

\subsection{Preliminary: Flow-Matching}

ZipVoice is built upon the flow-matching framework~\cite{lipman2023flow}, specifically conditional flow matching (CFM). We therefore begin with a brief introduction of CFM. 

CFM models learn to transform a simple initial distribution $p_0$ (e.g., standard Gaussian distribution) into a complex data distribution $p_1$ that approximates the real data distribution $q$. CFM model is parameterized as a time-dependent vector field $v_t(x_t; \theta), t\in[0,1]$, which can be used to construct a flow $\phi_t$ that transitions $p_0$ into $p_1$. Under the optimal transport form $\phi_t(x) = (1-t)x_0 + t x_1$, the CFM loss can be formulated as:

\begin{equation}
\label{equ:cfm}
L_{\text{CFM}} = E_{t, q(x_1), p_0(x_0)} || v_{t}( x_t; \theta) - (x_1-x_0)||^2
\end{equation}
where $x_t = (1-t)x_0 + t x_1$, $x_0$ is the Gaussian noise, $x_1$ is the data sample.

After training with the CFM loss, we can generate samples by solving the ordinary differential equation (ODE). Using the Euler solver as an example, starting from an initial sample $ x_0 \sim p_0$ (e.g., Gaussian noise), we can iteratively updates the sample towards the target distribution $ p_1 $.
For a discrete time sequence $ 0 = t_0  < \dots < t_k < \dots < t_K=1 
$, the Euler solver updates the sample at each step $ k $ as:
\begin{equation}
x_{t_{k+1}} = x_{t_k} + (t_{k+1} - t_k) \cdot v_{t_k}(x_{t_k}; \theta).
\end{equation}
This iteratively transforms $ x_0 $ into $ x_1 \sim p_1 $. The number of function evaluation (NFE) $K$ controls the trade-off between inference speed and sample quality.

Classifier-free guidance (CFG)~\cite{ho2021classifierfree} is widely used to improve the generation quality of flow-matching models. CFG requires dropping the condition with a certain probability during training, and performing linear interpolation of conditional and unconditional predictions during inference:

\begin{equation}
\label{equ:cfg}
\begin{split}
\tilde{v}_{t}(x_t, c, \omega; \theta) = & (1 + \omega)v_{t}(x_t, c; \theta) \\
& - \omega v_{t}(x_t, \emptyset; \theta)
\end{split}
\end{equation}
where $c$ and $\emptyset$ are condition and zero condition, $\omega$ is the CFG strength that controls the balance between fidelity and diversity.

\subsection{Overview}

The architecture of the ZipVoice model is illustrated in \autoref{fig:framework}. It consists of two components: a text encoder and a vector field estimator, both employing Zipformer as the backbone. We explain the advantage of this design in \autoref{sec:zipformer}.

The text sequence to be synthesized is first tokenized into text tokens, denoted as $y=(y_1, y_2, \ldots, y_N)$, where $y_i$ represents the $i$-th text token and $N$ is the token length. These text tokens are fed into the text encoder and transformed into text features $\hat{y} \in \mathbb{R}^{F \times N}$, where $F$ denotes the text feature dimension. The corresponding speech feature is denoted as $x_1 \in \mathbb{R}^{D \times T}$, where $D$ is the feature dimension and $T$ is the feature length. Then, the average upsampling (detailed in \autoref{sec:average_upsample}) is applied on the text feature to form the text condition $z \in \mathbb{R}^{F \times T}$.

To enable zero-shot TTS capabilities, we adopt the speech infilling task following \cite{le2023voicebox}. Specifically, a binary temporal mask $m \in \{0, 1\}^{D \times T}$, where 1 denotes masked positions, is applied to the speech feature $x_1 \in \mathbb{R}^{D \times T}$. The model is trained to reconstruct $m \odot x_1$ given the speech condition $(1-m) \odot x_1$, the interpolated noisy speech feature $x_t = (1-t)x_0 + tx_1$, and the text condition $z$. As shown in \autoref{fig:framework}, these three inputs have the same length and are concatenated along the feature dimension to form the input to the vector field estimator.

Then, the CFM objective in \autoref{equ:cfm} can be reformulated as:

\begin{equation}
\label{equ:cfm_tts}
\begin{aligned}
L_{\text{CFM-TTS}} = E_{t, q(x_1), p_0(x_0)} \big\lVert  \big( v_{t}( x_t, z, (1-m) \odot x_1 ; \theta) \\
{} - (x_1-x_0) \big) \odot m \big\rVert^2
\end{aligned}
\end{equation}
where we mask the training loss with $m$ to discard the losses at the positions of speech conditions.

The standard version of ZipVoice is trained with the above flow-matching loss. We also introduce a speedup version of ZipVoice that is further fine-tuned with flow distillation (detailed in \autoref{sec:distillation}).

After training, we use ODE solver to generate speech features, which is then transformed into waveforms with an additional vocoder model. We also introduce some inference-time strategies to improve the speech quality, which are described in \autoref{sec:inference}. 

\subsection{ZipVoice Architecture with Zipformer Backbone}
\label{sec:zipformer}

ZipVoice comprises a Zipformer-based text encoder and a Zipformer-based vector field estimator, with the latter housing the majority of the model's parameters. 
Zipformer is an encoder structure designed for automatic speech recognition (ASR). It distinguishes itself from the standard Transformer structures through three key designs: U-Net-like architecture, convolution modules, and attention weight reuse mechanism.

Zipformer is well-suited as the vector field estimator backbone for the following reasons. Firstly, the U-Net architecture, which processes feature representations at varying resolutions, is widely recognized as an effective inductive bias in diffusion models~\cite{si2024freeu,tianu}. Secondly, convolutional neural networks (CNNs) are proficient in capturing fine-grained local feature patterns, complementing the Transformer's strength in modeling long-range global dependencies. In TTS tasks, the nature of speech leads to strong correlations between adjacent hidden states~\cite{ren2019fastspeech,gulati2020conformer}, making CNNs particularly well-suited for enhancing modeling capacity.
Finally, Zipformer improves parameter efficiency by reusing attention weights across three modules within a single layer: two self-attention modules and a non-linear attention (NLA) module. This design not only reduces parameter count by sharing projection layers of query and key, but also enhances computational efficiency via attention weight sharing. Therefore, while originally engineered for ASR, these attributes endow Zipformer with ideal compatibility as the backbone of vector field estimators.

Although some recent flow-matching models~\cite{eskimez2024e2,chen2024f5} omit dedicated text encoders by feeding text embeddings directly into the vector field estimator, we find that a well-designed text encoder can improve speech intelligibility. Our Zipformer-based text encoder retains convolutional modules and attention weight reuse while omitting the U-Net structure, aligning with hybrid CNN-Transformer designs in prior works~\cite{kim2020glow,miao2020flow}.

\subsection{Speech-Text Alignment with Average Upsampling}
\label{sec:average_upsample}

A core challenge in NAR-TTS models is how to handle the length discrepancy between text and speech features, i.e., speech-text alignment. NAR-TTS models traditionally require explicit speech-text alignment during training (e.g., ASR-based forced alignment~\cite{le2023voicebox} or monotonic alignment search~\cite{kim2020glow}), along with a duration prediction model for inference.
This practice complicates the training and can degrade speech naturalness due to inaccurate duration estimates~\cite{anastassiou2024seed}.

Some recent NAR-TTS models~\cite{eskimez2024e2,simplespeech,lee2025dittotts,wang2025maskgct} eliminate the need for explicit alignment. A notable example is E2-TTS~\cite{eskimez2024e2}, which pads text tokens with filter tokens to match the speech length, allowing the model to learn alignment implicitly. This approach significantly simplifies the NAR-TTS architecture by removing the dependency on explicit alignment.

Despite its simplicity, this approach suffers from inaccurate alignment and slow convergence ~\cite{eskimez2024e2}. F5-TTS addresses this by incorporating a ConvNeXt~\cite{woo2023convnext} module to refine the padded text condition. ZipVoice instead applies a parameter-free average upsampling strategy to address this issue, under the intuitive assumption of uniform token durations, i.e., each token has identical duration within a sentence. Specifically, when there are $N$ text tokens and $T$ frames speech features frames, the duration of each text token is computed with:

\begin{equation}
d = \left\lfloor \frac{T}{N} \right\rfloor
\end{equation}
where the floor operation $\lfloor \cdot \rfloor$ ensures the expanded text features do not exceed the speech feature length. Under the practically valid assumption $T \geq N$, the minimum token duration is 1.

Each text embedding is repeated $d$ times and the length of the text feature is expanded from $N$ to $d \cdot N$. If $T > d \cdot N$, the expanded text features is further padded with $T - d \cdot N$ filler embeddings. The final text feature $z \in \mathbb{R}^{F \times T}$ is used as the text condition of the vector field estimator.

While this uniform-duration assumption is theoretically simplistic and has a considerable gap between real durations, it provides a reasonable initial text condition for the flow-matching model. Empirically, this straightforward strategy significantly improves alignment accuracy.

\subsection{Speedup ZipVoice with Flow Distillation}
\label{sec:distillation}

In this section, we introduce ZipVoice-Distill, a faster variant of ZipVoice obtained by distilling the ZipVoice model with a flow distillation method. Specifically, we construct a teacher vector field using two-step inference of the teacher model, with CFG applied at each step, then regress the student model's predictions onto this vector field. This enables the student to match the teacher's performance with fewer NFEs and no additional CFG-related evaluations. The flow distillation method is formally described below:

Given a pre-trained TTS model $\theta^T$ as the teacher, we initialize the student model $\theta^S$ with the parameters of $\theta^T$. 
Since we want the student model to benefit from the CFG inference while avoiding the additional model evaluation of CFG, we modify the student model to be conditioned on the CFG strength~\cite{meng2023distillation}. Specifically, the CFG strength $\omega$ is first processed through a Fourier embedding and a linear layer, and then integrated into the model, which is similar to how the time-step is incorporated into the vector field estimator.

For any time-step $t$ and input noisy speech $x_t$, we take two steps with the teacher model to reach a middle time-step $t_\text{mid}$ and a destination time-step $t_\text{dest}$:

\begin{equation}
\label{equ:teacher}
\begin{aligned}
x_{t_{\text{mid}}} &= \Phi(x_t, t, t_{\text{mid}}, c, \omega; \theta^{T}) \\
x_{t_{\text{dest}}} &= \Phi(x_{t_{\text{mid}}}, t_{\text{mid}}, t_{\text{dest}}, c, \omega; \theta^{T})
\end{aligned}
\end{equation}
where the one-step ODE solver $\Phi$ is defined as:
\begin{equation}
\Phi(x_t, t, t_{\text{mid}}, c, \omega; \theta^{T}) = x_t + (t_{\text{mid}} - t) \tilde{v}_{t}(x_t, c, \omega; \theta^{T})
\end{equation}
where $\tilde{v}_{t}(x_t, c, \omega; \theta^{T})$ is the prediction with CFG as in \autoref{equ:cfg}.

The two step sizes, $t_{\text{mid}} -t $ and $t_{\text{dest}} - t_{\text{mid}}$ are uniformly sampled from $[0, \Delta t_\text{max}]$. 
$\omega$ is also uniformly sampled between a given interval $[\omega_\text{min}, \omega_\text{max}]$. We empirically find that dynamically sampling these values yields better performance than using fixed values.

Then, the teacher vector field is computed as:
\begin{equation}
    v^{\text{T}} = \frac{x_{t_{\text{dest}}} - x_t}{t_{\text{dest}} - t}
\end{equation}

Finally, the flow distillation loss is computed as:

\begin{equation}
\label{equ:flow_distillation}
\begin{aligned}
L_{\text{FD}} = E_{t, q(x_1), p_0(x_0)} \big\lVert \big( v_{t}( x_t, \hat{e}, (1-m) \odot x_1 ; \theta) \\
{} - v^{T} \big) \odot m \big\rVert^2
\end{aligned}
\end{equation}

Since the teacher prediction in \autoref{equ:teacher} is made with CFG, after flow distillation, the student model can benefit from CFG by simply passing the CFG strength $\omega$ as one input of the TTS model, avoiding doubled model evaluation in each step.

After completing flow distillation using the fixed teacher model $\theta^{T}$, we can conduct a second distillation phase starting from the latest student mode $\theta^{S}$. The key distinction lies in the construction of the teacher vector field, which now employs the student model's exponential moving averaging (EMA) version  $\tilde{\theta}$, which is updated with:
\begin{equation}
\tilde{\theta^{S}}= (1-\beta)\theta^{S} + \beta \tilde{\theta}
\end{equation}
after each training update, where $\beta$ is the EMA decay factor.

This teacher vector field derived from the continuously evolving student model enables the second distillation phase to refine the student's performance iteratively.

\subsection{Inference Strategy}
\label{sec:inference}

As a zero-shot TTS model, in addition to the text $y^{synthesis}$ to be synthesized, ZipVoice also requires an audio prompt $s^{prompt}$ and its transcription $c^{prompt}$ mimic its voice. 

The sentence duration of the synthesized audio is estimated based on the token length ratio between prompt transcription and the text to be synthesized:

\begin{equation}
    T^{synthesis} = T^{prompt} \cdot \frac{|y^{synthesis}|}{|y^{prompt}|}
\end{equation}
where $T^{prompt}$ is sentence duration of the prompt audio.

The input of the text encoder is the concatenation of tokenized text tokens $y^{synthesis}$ and $y^{prompt}$. Then, the text condition is constructed with the average upsampled text features. The audio condition is formed by padding the audio prompt to the length $T^{prompt} + T^{synthesis}$. The initial noisy speech is sampled from a standard Gaussian distribution. Finally, the synthesized speech is sampled with an ODE solve.

We use a time-dependent CFG strategy to achieve a better trade-off between speaker similarity and intelligibility: in early NFEs, only the text condition is dropped for unconditional predictions, while in later steps, both text and audio conditions are dropped.

\begin{table*}[!ht]
    \centering
    \caption{Evaluation results of zero-shot TTS models trained on large-scale datasets. The boldface denotes the best result. $*$ Results from papers. $\S$ Unofficial implementations. $\dagger$ Inference with official checkpoints. $\ddagger$ trained by us with official codes.}
    \label{tab:large_scale_results}
    \resizebox{\textwidth}{!}{
    \begin{tabular}{lccc*{3}{c}*{3}{c}*{3}{c}cc} 
        \toprule
        \multirow{3}{*}{\textbf{Model}} & \multirow{3}{*}{\textbf{Data (hrs)}} & \multirow{3}{*}{\textbf{Params}} & 
        \multicolumn{9}{c}{\textbf{Objective Metrics}} &
        \multicolumn{2}{c}{\multirow{2}{*}{\textbf{Subjective Metrics}}} \\ 
        \cmidrule(lr){4-12} 
        & & & \multicolumn{3}{c}{\textbf{LibriSpeech-PC test-clean}} & 
        \multicolumn{3}{c}{\textbf{Seed-TTS test-en}} & 
        \multicolumn{3}{c}{\textbf{Seed-TTS test-zh}} &
        \multicolumn{2}{c}{} \\ 
        \cmidrule(lr){4-6} \cmidrule(lr){7-9} \cmidrule(l){10-12} \cmidrule(l){13-14}
        & & & \textbf{SIM-o $\uparrow$} & \textbf{WER $\downarrow$} & \textbf{UTMOS $\uparrow$} & \textbf{SIM-o $\uparrow$} & \textbf{WER $\downarrow$} & \textbf{UTMOS $\uparrow$} & \textbf{SIM-o $\uparrow$} & \textbf{WER $\downarrow$} & \textbf{UTMOS $\uparrow$} & \textbf{CMOS $\uparrow$} & \textbf{SMOS $\uparrow$} \\
        \midrule
        Ground Truth & - & - & 0.690 & 1.87 & 4.10 & 0.734 & 2.14 & 3.52 & 0.755 & 1.25 & 2.78 & 0 & 3.36 \\
        \midrule
        \multicolumn{14}{l}{\textit{\textbf{AR Models}}}\\
        CosyVoice$^*$ & 170K Multi. & 416M & - & - & - & 0.609 & 4.29 & - & 0.723 & 3.63 & - & - & - \\
        CosyVoice 2$^*$ & 167K Multi. & 618M & - & - & - & 0.652 & 2.57 & - & 0.748 & 1.45 & - & - & - \\
        Spark-TTS$^*$ & 102K Multi. & 507M & - & - & - & 0.584 & 1.98 & - & 0.672 & \textbf{1.20} & - & - & - \\
        \midrule
        \multicolumn{14}{l}{\textit{\textbf{NAR Models}}}\\
        MaskGCT$^\dagger$ & 100K Emilia & 1048M & 0.691 & 2.26 & 3.91 & \textbf{0.713} & 2.88 & 3.55 & \textbf{0.773} & 2.40 & 2.63 & -0.08 & \textbf{4.10} \\
        E2-TTS$^\S$ (32 NFE) & 100K Emilia & 333M & \textbf{0.700} & 2.49 & 3.47 & 0.706 & 2.32 & 3.21 & 0.713 & 1.91 & 2.26 & - & - \\
        F5-TTS$^\dagger$ (32 NFE) & 100K Emilia & 336M & 0.655 & 1.89 & 3.89 & 0.664 & 1.85 & 3.72 & 0.750 & 1.53 & 2.93 & -0.03 & 3.76 \\
        F5-TTS$^\ddagger$ (32 NFE) & 100K Emilia & 155M & 0.615 & 2.10 & 3.84 & 0.628 & 1.96 & 3.66 & 0.733 & 1.57 & 2.93 & - & - \\
        \addlinespace

        ZipVoice (16 NFE) & 100K Emilia & 123M & 0.668 & 1.64 & 3.98 & 0.697 & 1.70 & 3.82 & 0.751 & 1.40 & 3.15 & \textbf{0.17} & 3.94 \\
        ZipVoice-Distill (8 NFE) & 100K Emilia & 123M & 0.647 & 1.54 & \textbf{4.11} & 0.670 & 1.62 & \textbf{3.91} & 0.740 & 1.34 & \textbf{3.18} & 0.16 & 3.88 \\
        ZipVoice-Distill (4 NFE) & 100K Emilia & 123M & 0.657 & \textbf{1.51} & 4.05 & 0.679 & \textbf{1.64} & 3.91 & 0.748 & 1.39 & 3.16 & 0.05 & 3.84 \\
        \bottomrule
    \end{tabular}
    }
\end{table*}

\section{Related Works}

Accelerating TTS models has long been a critical challenge. Early end-to-end TTS models~\cite{wang2017tacotron,shen2018natural} suffered from slow inference due to AR speech generation. FastSpeech~\cite{ren2019fastspeech} addressed this with NAR architectures. But the limited modeling capacity in early NAR models often resulted in blurry or oversmoothed speech~\cite{ren2022revisiting}. Diffusion models~\cite{popov2021grad} can improve speech quality but rely on many NFEs, making them inefficient. Flow-matching~\cite{lipman2023flow} has since been adopted in various works~\cite{mehta2024matcha,le2023voicebox} to decrease NFEs. Furthermore, consistency distillation~\cite{song2023consistency}, consistency training~\cite{song2023consistency} and ReFlow~\cite{liuflow} have also been adopted in some works~\cite{ye2023comospeech,guan2024reflow,guo2024voiceflow,ye2024flashspeech,wang2025slimspeech} to decrease NFEs with a second-stage training or auxiliary training objectives. Parallel to the above efforts, some works~\cite{luo2021lightspeech,mehta2024matcha} accelerate TTS models with efficient model structures.
However, most related works focus on small datasets and fail to match SOTA performance across metrics. Our work shows a significant speedup while maintaining SOTA performance across metrics. Moreover, we focus on speeding up NAR-TTS models without explicit speech-text alignment, which is an inherently more challenging and under-explored problem.

\section{Experimental Setup}

\subsection{Dataset}
We train the ZipVoice model on the large-scale 100k hours Emilia~\cite{he2024emilia} dataset for comparison with existing SOTA zero-shot TTS models. We also train on the small-scale 585 hours LibriTTS~\cite{zen2019libritts} dataset for development and ablation experiments. 
We evaluated our zero-shot TTS model on three widely adopted benchmarks: LibriSpeech-PC~\cite{meister2023librispeech} test-clean subset~\cite{chen2024f5} with 1127 English samples, Seed-TTS test-en~\cite{anastassiou2024seed} with 1088 English samples from Common Voice~\cite{ardila2020common}, and Seed-TTS test-zh with 2020 Chinese samples from DiDiSpeech~\cite{guo2021didispeech}. 

\subsection{Model}
ZipVoice consists of a Zipformer-based text encoder and a Zipformer-based vector field estimator. The text encoder consists of 4 Zipformer layers, each having an encoder dimension of 192, a feedforward dimension of 512. For the vector field estimator, it contains 5 stacks of Zipformer encoder layers operating at $[1 \times, 2\times, 4\times, 2\times, 1\times]$ downsampling rates and with $[2, 2, 4, 4, 4]$ layers, respectively. Each layer in the vector field estimator has a embedding dimension of 512 and a feedforward dimension of 1536. The total number of parameters of ZipVoice models are around 123M.

\subsection{Training}

On Emilia/LibriTTS datasets, ZipVoice is trained for 1M/60k updates with the flow-matching objective, followed by 62k/12k updates with the flow distillation objective, employing total batch sizes of 4k/2k seconds. During the training with flow-matching objective, the text condition is dropped with a probability of 20\% for CFG. For the speech infilling task, masking lengths are randomly sampled between 70\% and 100\% of the speech feature length. We use phoneme tokens for Emilia and characters tokens for LibriTTS.

\subsection{Inference}

We use Euler ODE solver for sampling. The pre-trained Vocos~\cite{siuzdak2024vocos} vocoder, trained on the LibriTTS dataset, is used to convert the generated speech features into waveforms.

\subsection{Metrics}

Three model-based reproducible metrics are used for subjective evaluation. For intelligibility, we use the word error rate (WER) between the transcription of synthesized speech and the input text. We use the Whisper-large-v3 ASR model~\cite{radford2023robust} for Seed-TTS test-en, Paraformer-zh~\cite{gao2022paraformer} for Seed-TTS test-zh dataset, and the Hubert-based ASR model~\cite{hsu2021hubert} for LibriSpeech-PC test-clean. For speaker similarity, we use speaker embeddings from a WavLM-based~\cite{chen2022wavlm} ECAPA-TDNN model~\cite{desplanques2020ecapa} to measure the cosine similarity between the original prompt speech and synthesized speech, denoted as SIM-o~\cite{le2023voicebox}. For naturalness, we use a neural network-based mean opinion score (MOS) prediction model UTMOS~\cite{saeki2022utmos}.

As for objective evaluation metrics, we use Comparative Mean Opinion Scores (CMOS) and Similarity Mean Opinion Scores (SMOS) to evaluate comparative quality and speaker similarity of the synthesized speech.

\section{Experimental Results}

\subsection{Comparison with Baseline Models on Large-Scale Datasets}

In \autoref{tab:large_scale_results}, we compare ZipVoice with existing SOTA zero-shot TTS models trained on large-scale datasets. Specifically, we benchmark against three AR models (Cosyvoice~\cite{du2024cosyvoice}, Cosyvoice 2~\cite{du2024cosyvoice2}, and Spark-TTS~\cite{wang2025spark}) and three NAR model (MaskGCT~\cite{wang2025maskgct}, E2-TTS~\cite{eskimez2024e2}, and F5-TTS~\cite{chen2024f5}). For AR models, we report results from \cite{wang2025spark}. For NAR models, we use the unofficial implementation in \cite{chen2024f5} for E2-TTS and official implementations for other models. Additionally, we trained a smaller version of F5-TTS with its official codes to illustrate performance degradation due to model size reduction.

Despite its compact model size, ZipVoice delivers performance comparable to other SOTA models that have substantially more parameters. ZipVoice demonstrates clear advantages in WER and UTMOS, while remaining comparable in SIM-o. Although not achieving the best result on all metrics, its overall performance is highly competitive.
Comparing ZipVoice and ZipVoice-Distill, flow distillation allows ZipVoice-Distill to prioritize speech quality (WER and UTMOS) with a marginal reduction in speaker similarity (SIM-o).

\begin{table}[!ht]
    \centering
    \caption{Inference speed comparison of models.}
    \label{tab:inference_speed}
    \resizebox{0.7\columnwidth}{!}{
    \begin{tabular}{lcccc}
        \toprule
        \multirow{2}{*}{\textbf{Model}} & \multirow{2}{*}{\textbf{Params}} & \multicolumn{2}{c}{\textbf{RTF}$\downarrow$} \\
        \cmidrule(lr){3-4}
        & & \textbf{GPU} & \textbf{CPU} \\
        \midrule
        F5-TTS (32 NFE) & 336M& 0.2958 & 37.284 \\
        ZipVoice (16 NFE) & 123M & 0.0557 & 9.5529 \\
        ZipVoice-Distill (8 NFE) & 123M & 0.0233 & 2.4177 \\
        ZipVoice-Distill (4 NFE) & 123M & \textbf{0.0125} & \textbf{1.2202} \\
        \bottomrule
    \end{tabular}
    }
\end{table}

Since F5-TTS is the fastest model in \autoref{tab:large_scale_results}, we compare the real-time factor (RTF) of F5-TTS and ZipVoice in \autoref{tab:inference_speed} with same devices and PyTorch versions. All models use the same Vocos~\cite{siuzdak2024vocos} vocder, whose RTF is negligible compared with the TTS models. We use 3s prompt speech to generate 10s speech. ZipVoice-Distill (4 NFE) is 23.7 times faster than F5-TTS on the NVIDIA H20 GPU, and 32.6 times faster on a single thread of an Intel(R) Xeon(R) Platinum 8457C CPU. The speed advantage of ZipVoice-Distill comes from three aspects: compact model size, fewer NFEs, and avoiding the additional model inference introduced by CFG. ZipVoice-Distill is close to real-time on a single CPU thread, which can potentially expand the applicable devices of SOTA zero-shot TTS models.

\subsection{Comparison with Baseline Models on Small-Scale Datasets}

We also train a model on a LibriTTS, a small-scale dataset, and compare it with other models trained on small-scale datasets. Specifically, we compare with 3 baseline models: VALL-E~\cite{chen2025neural}, YourTTS~\cite{casanova2022yourtts} and F5-TTS. We use the unofficial implementation in \cite{zhang2024amphion} for VALL-E, and use the official implementation for YourTTS. We train a F5-TTS model on LibriTTS with their official codes for 500k updates.

\begin{table}[!ht]
    \centering
    \caption{Evaluation results of zero-shot TTS models trained on small-scale datasets. The boldface denotes the best result. $\S$ Unofficial implementations. $\dagger$ Inference with official checkpoints. $\ddagger$ trained by us with official codes.}
    \label{tab:small_scale_results}
    \resizebox{\columnwidth}{!}{
    \begin{tabular}{lccccc}
        \toprule
        \multirow{2}{*}{\textbf{Model}} & \multirow{2}{*}{\textbf{Data (hrs)}} & \multirow{2}{*}{\textbf{Params}} & 
        \multicolumn{3}{c}{\textbf{LibriSpeech-PC test-clean}} \\
        \cmidrule(lr){4-6}
        & & & \textbf{SIM-o$\uparrow$} & \textbf{WER$\downarrow$} & \textbf{UTMOS$\uparrow$}  \\
        \midrule
        Ground-truth & -- & -- & 0.690 & 1.87 & 4.10 \\
        \midrule
        VALL-E$^\S$ & 6K Libri-light & 367M & 0.369 & 6.20 & 3.01  \\
        YourTTS$^\dagger$ & 474 Multi. & 87M & 0.462 & 6.37 & 3.68 \\ 
        F5-TTS (32 NFE)$^\ddagger$ & 555 LibriTTS & 158M & 0.584 & 1.78 & 4.14 \\
        \midrule
        ZipVoice (8 NFE) & 555 LibriTTS & 123M & \textbf{0.610} & 1.69 & 4.16 \\ 
        ZipVoice-Distill (4 NFE) & 555 LibriTTS & 123M & 0.606 & \textbf{1.68} & \textbf{4.22} \\ 
        \bottomrule
    \end{tabular}
    }
\end{table}

As shown in \autoref{tab:small_scale_results}, in the small-scale data setting, ZipVoice still shows competitive performance. Among the three baselines, the flow-matching-based model F5-TTS demonstrates the strongest performance, demonstrating the robustness of flow-matching model. Although with smaller model size and fewer training updates, ZipVoice and ZipVoice-Distill are consistently better than F5-TTS on all metrics.

\subsection{Ablation Experiments on the ZipVoice Structure}

\begin{table}[!ht]
    \centering
    \caption{Ablation studies on the ZipVoice architecture.}
    \label{tab:zipvoice_structure_ablation}
    \resizebox{0.7\columnwidth}{!}{
    \begin{tabular}{lccc}
        \toprule
        \multirow{2}{*}{\textbf{Model architecture}} & 
        \multicolumn{3}{c}{\textbf{LibriSpeech-PC test-clean}} \\
        \cmidrule(lr){2-4}
        & \textbf{SIM-o}$\uparrow$ & \textbf{WER}$\downarrow$ & \textbf{UTMOS}$\uparrow$  \\
        \midrule
        ZipVoice & 0.610 & 1.69 & 4.16  \\ 
        \midrule
        w/o text encoder & 0.597 & 2.04 & 4.15  \\ 
        w/o average upsample & 0.513 & 20.19 & 3.89 \\
        \quad + w/ ConvNext & 0.524 & 15.49 & 3.80  \\ 
        \bottomrule
    \end{tabular}
    }
\end{table}

In this section, we show the impact of two architecture designs in ZipVoice: the text encoder and the average upsampling strategy. As shown in \autoref{tab:zipvoice_structure_ablation}, both designs are important in ensuring high speech intelligibility (low WER). Notably, removing the average upsampling leads to a substantial performance degradation. We also tested the ConvNext-based text condition refinement method proposed in \cite{chen2024f5}. While the method also reduces WER to some extent, the average upsampling strategy still shows a clear performance advantage.

\subsection{Ablation Experiments on the Zipformer Backbone}

\begin{table}[!ht]
    \centering
    \caption{Ablation studies on the backbone structure.}
    \label{tab:zipformer_backbone_ablation}
    \resizebox{0.8\columnwidth}{!}{
    \begin{tabular}{lccc}
        \toprule
        \multirow{2}{*}{\textbf{Backbone Structure}} & 
        \multicolumn{3}{c}{\textbf{LibriSpeech-PC test-clean}} \\
        \cmidrule(lr){2-4}
        & \textbf{SIM-o}$\uparrow$ & \textbf{WER}$\downarrow$ & \textbf{UTMOS}$\uparrow$ \\
        \midrule
        Zipformer & 0.610 & 1.69 & 4.16 \\ 
        \midrule
        w/o convolution & 0.586 & 9.79 & 4.01 \\ 
        w/o downsampling & 0.557 & 3.70 & 4.02 \\ 
        w/o downsample \& bypass & 0.562 & 6.35 & 4.08 \\ 
        w/o bypass & 0.179 & 98.89 & 1.25 \\ 
        w/o NLA & 0.548 & 1.83 & 3.99 \\ 
        w/o share attention weight & 0.595 & 1.75 & 4.18 \\ 
        \bottomrule
    \end{tabular}
    }
\end{table}

We conduct detailed ablation experiments to validate the effectiveness of using Zipformer as the backbone of the vector field estimator. As shown in \autoref{tab:zipformer_backbone_ablation}, convolution modules are important in assuring high intelligibility. The U-Net-like downsampling and bypass are consistently helpful in all metrics, showing the effectiveness of the U-Net-like inductive bias on the flow-matching-based TTS model. Notably, removing downsampling but keeping the bypass leads to an architecture similar to the flat Unet-style linked Transformer architecture in \cite{eskimez2024e2}, which has better results than the model without both downsampling and bypass. And removing bypass connections while retaining downsampling leads to severely degraded results, highlighting the importance of cross-resolution bypass connections in U-net style downsampling structure.  Finally, we show the impact of re-using the attention weight. By re-using the attention weight in the NLA module, the performance is consistently improved in all metrics. And the design of sharing attention weight in two self-attention module leads to similar results with the baseline but has improved efficiency.

\subsection{Experiments with Different Distillation Methods}

\begin{table}[!ht]
    \centering
    \caption{Ablation study of distillation methods with different NFE. The boldface denotes best results in each NFE.}
    \label{tab:distillation_ablation}
    \resizebox{0.8\columnwidth}{!}{
    \begin{tabular}{lcccc}
        \toprule
        \multirow{2}{*}{\textbf{Distillation Method}} & 
        \multirow{2}{*}{\textbf{NFE}} & 
        \multicolumn{3}{c}{\textbf{LibriSpeech-PC test-clean}} \\
        \cmidrule(lr){3-5}
        & & \textbf{SIM-o$\uparrow$} & \textbf{WER$\downarrow$} & \textbf{UTMOS$\uparrow$} \\
        \midrule
        \multirow{3}{*}{Model w/o distillation} & 1 & 0.171 & 92.17 & 1.30  \\ 
        & 2 & 0.475 & 15.00 & 1.83 \\ 
        & 4 & \textbf{0.634} & 2.11 & 3.84 \\ 
        \midrule
        \multirow{3}{*}{Consistency distillation}  & 1 & 0.441 & 20.72 & 1.53 \\ 
        & 2 & 0.571 & 4.06 & 3.08 \\ 
        & 4 & 0.568 & 1.97 & 3.30 \\ 
        \midrule
        \multirow{3}{*}{ReFlow} & 1 & \textbf{0.512} & \textbf{17.40} & \textbf{1.90} \\ 
        & 2 & \textbf{0.601} & 5.36 & 3.39 \\ 
        & 4 & 0.608 & 2.56 & 3.92 \\ 
        \midrule
        \multirow{3}{*}{Our flow distillation} & 1 & 0.398 & 18.81 & 1.58 \\ 
        & 2 & 0.588 & \textbf{2.33} & \textbf{3.73} \\ 
        & 4 & 0.606 & \textbf{1.68} & \textbf{4.22} \\ 
        \bottomrule
    \end{tabular}
    }
\end{table}

In this section, we demonstrate the effectiveness of the proposed flow distillation method. We compare our flow distillation method with two popular acceleration methods of flow-matching models: consistency distillation~\cite{song2023consistency} and ReFlow~\cite{liuflow}. We train all models with the same number of updates. As shown in \autoref{tab:distillation_ablation}, consistency distillation effectively improves the performance at 1 and 2 NFEs, but fails to improve performance at 4 NFEs. However, achieving satisfactory performance with just 1 or 2 NFEs is challenging for ZipVoice due to the absence of explicit token-level duration in the text condition. Consequently, consistency distillation is not applicable in this context. ReFlow~\cite{liuflow} improves UTMOS across all NFE settings, but leads to WER degradation with with NFEs larger than 1. In contrast, our flow distillation method can consistently improves performance across different NFEs and achieves the best result at 4 NFEs.

\section{Conclusion}

In this paper, we propose ZipVoice, a fast and high-quality flow-matching-based zero-shot TTS model. Leveraging a Zipformer-based backbone for the vector field estimator, ZipVoice achieves parameter efficiency while maintaining strong modeling capabilities. An average upsampling strategy, combined with a Zipformer text encoder, ensures stable speech-text alignment and intelligibility. In addition, a flow distillation method is proposed to reduce the NFEs, significantly accelerating inference. Experimental results show that ZipVoice matches SOTA zero-shot TTS models in speech quality but with fewer parameters and faster inference speed.

\bibliographystyle{IEEEtran}
\bibliography{bib}

\end{document}